\documentclass[prd,twocolumn,groupedaddress,showpacs,nofootinbib]{revtex4}
\usepackage{graphicx}
\usepackage{amsmath}
\usepackage{epsfig}

\begin{document}

\title{Leptogenesis Bound on Spontaneous Symmetry Breaking of Global Lepton
Number}

\author{Pei-Hong Gu$^{1}_{}$}
\email{pgu@ictp.it}

\author{Utpal Sarkar$^{2}_{}$}
\email{utpal@prl.res.in}

\affiliation{ $^{1}_{}$The Abdus Salam International Centre for
Theoretical Physics, Strada Costiera 11, 34014 Trieste, Italy \\
$^{2}_{}$Physical Research Laboratory, Ahmedabad 380009, India}

\begin{abstract}

We propose a new class of leptogenesis bounds on the spontaneous
symmetry breaking of global lepton number. These models have a
generic feature of inducing new lepton number violating
interactions, due to the presence of the Majorons. We analyzed the
singlet Majoron model with right-handed neutrinos and find that the
lepton number should be broken above $10^{5}_{}\,\textrm{GeV}$ to
realize a successful leptogenesis because the annihilations of the
right-handed neutrinos into the massless Majorons and into the
standard model Higgs should go out of equilibrium before the
sphaleron process is over. We then argue that this type of
leptogenesis constraint should exist in the singlet-triplet Majoron
models as well as in a class of R-parity violating supersymmetric
Majoron models.

\end{abstract}


\maketitle

There have been many experiments establishing the tiny but nonzero
neutrino masses although we are yet to find out if the neutrinos are
Majorana or Dirac particles. If the neutrinos are Majorana
particles, lepton number is necessarily broken. In the simplest
extension of the standard model (SM) for the small neutrino masses,
the lepton number is explicitly broken by the supplement of a
Weinberg dimension-5 operator \cite{weinberg1979} in the Lagrangian.
This operator can be generated in the renormalizable seesaw
\cite{minkowski1977} scenario, where an explicit lepton number
breaking Majorana mass term for the right-handed neutrinos
\cite{minkowski1977} and(or) trilinear interaction for the triplet
and doublet Higgs scalars \cite{mw1980} are(is) introduced to the
SM.

The seesaw models also provide a solution to the puzzle of the
matter-antimatter asymmetry in the universe through leptogenesis
\cite{fy1986}: the CP-violating and out-of-equilibrium decays of the
right-handed neutrinos \cite{fy1986} and(or) triplet Higgs scalars
\cite{mz1992,ms1998,hs2003} to the SM particles can generate a
lepton asymmetry, which is partially converted to a baryon asymmetry
through the sphaleron \cite{krs1985} action and hence accounts for
the matter-antimatter asymmetry. For this purpose, the Yukawa
interactions of the right-handed neutrinos while the gauge, Yukawa
and trilinear interactions of the triplet Higgs scalars should
satisfy the out-of-equilibrium condition before the sphaleron
process becomes very weak at $T_{sph}\sim 100\,\textrm{GeV}$.

If we further consider that the original seesaw scheme to be
originating from a more fundamental theory, where the lepton number
is spontaneously broken locally or globally, there will be some new
particles coupling to the right-handed neutrinos and(or) triplet
Higgs scalars. This implies additional out-of-equilibrium conditions
for a successful leptogenesis, which, in turn, constrains the
breaking scale of the lepton number. For example, in the left-right
symmetric model, the right-handed gauge bosons should be heavier
than $10^{6-7}_{}\,\textrm{GeV}$ to guarantee the departure from
equilibrium of the right-handed neutrinos \cite{mz1992}.

In this note, we first study the constraint from leptogenesis on the
singlet Majoron model of Chikashige, Mohapatra and Peccei
\cite{cmp1980}. In this model, a singlet scalar, driving the
spontaneous breakdown of the global lepton number conservation, is
introduced to the SM in addition to the three right-handed
neutrinos. The right-handed neutrinos couple to the singlet scalar
and hence obtain a Majorana mass term after the lepton number is
broken. These couplings also result in some scattering processes for
the right-handed neutrinos to the light species. Therefore, the
breaking scale of the lepton number conservation should be under the
authority of the successful leptogenesis. We then extend this
analysis to the singlet-triplet Majoron model and the R-parity
violating supersymmetric Majoron model, where this new leptogenesis
constraint is also applicable for certain range of the parameters.

We start with the singlet Majoron model. In this model, the kinetic
and Yukawa terms of the lepton sector would be
\begin{eqnarray} \label{kinetic}
\mathcal{L}_{K}^{}&\supset&
\overline{\psi_L^{}}i\gamma^\mu_{}D_\mu^{}\psi_L^{}+\overline{\ell_R^{}}i\gamma^\mu_{}D_\mu^{}\ell_R^{}
+\overline{N_R^{}}i\gamma^\mu_{}\partial_\mu^{}
N_R^{}\nonumber\\
&&
+\left(\partial_\mu^{}\chi\right)^\dagger_{}\left(\partial^\mu_{}\chi\right)\,\\
\label{yukawa} \mathcal{L}_{Y}^{}&\supset& -
y_{\ell}^{}\overline{\psi_{L}^{}}\tilde{\phi}
\ell_{R}^{}-y_{\nu}^{}\overline{\psi_{L}^{}}\phi
N_{R}^{}-\frac{1}{2}h\chi\overline{N_{R}^{c}}N_{R}^{}+\textrm{H.c.}\,.
\end{eqnarray}
Here $\psi_{L}^{}(\mathbf{1},\mathbf{2},-\frac{1}{2})$ and
$\ell_{R}^{}(\mathbf{1},\mathbf{1},-1)$, respectively, are the SM
lepton doublets and singlets,
$\phi(\mathbf{1},\mathbf{2},-\frac{1}{2})$ is the SM Higgs doublet,
$N_{R}^{}(\mathbf{1},\mathbf{1},0)$ denotes the right-handed
neutrinos, $\chi(\mathbf{1},\mathbf{1},0)$ is the Higgs singlet,
where the transformations are given under the SM gauge group
$SU(3)_{c}^{}\times SU(2)_L^{} \times U(1)_{Y}^{}$. Conventionally,
we assign the Higgs singlet $\chi$ a lepton number $L=-2$, as the
right-handed neutrinos $N_R^{}$ have $L=+1$. As a result, the Yukawa
interaction (\ref{yukawa}) is lepton number conserving since $L=1$
for the SM leptons $\psi_{L_\alpha}^{}$ and $\ell_R^{}$ while $L=0$
for the SM Higgs $\phi$. The general scalar potential contains the
quadratic and quartic terms as below,
\begin{eqnarray}
\label{potential}
V(\chi,\phi)&=&-\mu_{1}^{2}\left(\chi^{\dagger}_{}\chi\right)
+\lambda_{1}^{}\left(\chi^{\dagger}_{}\chi\right)^{2}_{}-
\mu_{2}^{2}\left(\phi^{\dagger}_{}\phi\right)\nonumber\\
&&+\lambda_{2}^{}(\phi^{\dagger}_{}\phi)^{2}_{}
+\lambda_{3}^{}\left(\chi^{\dagger}_{}\chi\right)\left(\phi^{\dagger}_{}\phi\right)\,,
\end{eqnarray}
where
$\lambda_3^{}>-2\left(\lambda_1^{}\lambda_2^{}\right)^{\frac{1}{2}}_{}$
so that the potential is bounded from below.

\begin{figure*}
\vspace{5.0cm} \epsfig{file=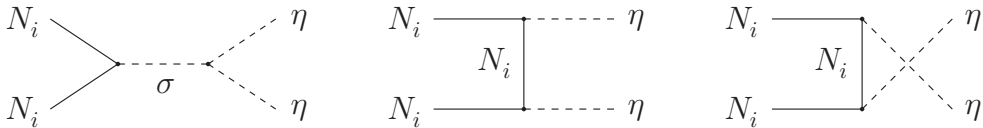, bbllx=5.3cm, bblly=6.0cm,
bburx=15.3cm, bbury=16cm, width=8.5cm, height=8.5cm, angle=0,
clip=0} \vspace{-10cm} \caption{\label{annihilation1} The pair
annihilation of the Majorana neutrinos into the
Majorons.}\vspace{4.5cm}
\end{figure*}

\begin{figure}
\vspace{4.5cm} \epsfig{file=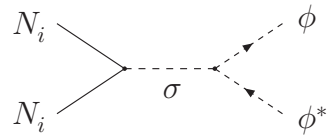, bbllx=0.3cm, bblly=6.0cm,
bburx=10.3cm, bbury=16cm, width=8.5cm, height=8.5cm, angle=0,
clip=0} \vspace{-10cm} \caption{\label{annihilation2} The pair
annihilation of the Majorana neutrinos into the pair of the SM
Higgs.}
\end{figure}

Once the Higgs singlet $\chi$ develops its vacuum expectation value
(VEV), we can write
\begin{eqnarray}
\label{sigmaeta}
\chi=\frac{1}{\sqrt{2}}(f+\sigma)e^{i\frac{\eta}{f}}_{}\,,
\end{eqnarray}
where
$\displaystyle{f=(\mu_{1}^{2}/\lambda_{1}^{})^{\frac{1}{2}}_{}}$ is
the VEV, $\sigma$ is the physical Higgs boson with the mass
\begin{eqnarray}
M_{\sigma}^{}=(2\lambda_{1}^{})^{\frac{1}{2}}_{}f\,,
\end{eqnarray}
while $\eta$ is the massless Majoron. By taking the phase rotations,
\begin{subequations}
\label{new}
\begin{eqnarray}
e^{i\frac{\eta}{2f}}_{}N_R^{}&\rightarrow& N_R^{}\,,\\
e^{i\frac{\eta}{2f}}_{}\psi_L^{}&\rightarrow& \psi_L^{}\,,\\
e^{i\frac{\eta}{2f}}_{}\ell_R^{}&\rightarrow& \ell_R^{}\,,
\end{eqnarray}
\end{subequations}
we can perform the following interactions,
\begin{eqnarray}
\label{interactions}
\mathcal{L}&\supset&\frac{1}{2f}\left(\overline{\psi_L^{}}\gamma^\mu_{}\psi_L^{}
+\overline{\ell_R^{}}\gamma^\mu_{}\ell_R^{}+\overline{N_R^{}}\gamma^\mu_{}N_R^{}
\right)\partial_{\mu}^{}\eta\nonumber\\
&&+\frac{1}{f}\sigma\partial^{\mu}_{}\eta\partial_{\mu}^{}\eta-
\left[y_{\ell}^{}\overline{\psi_{L}^{}}\tilde{\phi}
\ell_{R}^{}+y_{\nu}^{}\overline{\psi_{L}^{}}\phi
N_{R}^{}\right.\nonumber\\
&&\left.+\frac{1}{2\sqrt{2}}h\left(f+\sigma\right)\overline{N_{R}^{c}}N_{R}^{}
+\textrm{H.c.}\right]-\lambda_{3}^{}f\sigma\phi^\dagger_{}\phi \,.
\end{eqnarray}
For convenience, we choose the basis in which the Yukawa couplings
$h\rightarrow \hat{h}$ are real and diagonal by the proper phase
rotation and then define the Majorana neutrinos,
\begin{eqnarray}
N_{i}^{}=N_{R_i^{}}^{}+N_{R_i^{}}^{c}\,,
\end{eqnarray}
with the masses,
\begin{eqnarray}
M_{N_i^{}}^{}=\frac{1}{\sqrt{2}}\hat{h}_{i}^{}f\,.
\end{eqnarray}
We then obtain
\begin{eqnarray} \label{leptogenesis}
\mathcal{L}&\supset& - \left(y_{\nu}^{}\overline{\psi_{L}^{}}\phi
N+\textrm{H.c.}\right)-\frac{1}{2}M_{N}^{}\overline{N}N\nonumber\\
&&+\frac{1}{4f}\overline{N}\gamma^\mu_{}\gamma_5^{}N\partial_{\mu}^{}\eta
-\frac{1}{2\sqrt{2}}\hat{h}\sigma\overline{N}N\nonumber\\
&&+\frac{1}{f}\sigma
\partial^{\mu}_{}\eta \partial_{\mu}^{}\eta-\lambda_{3}^{}f\sigma\phi^\dagger_{}\phi \,.
\end{eqnarray}
Here the first line definitely indicates the conventional
leptogenesis scenario in the seesaw context, the second and third
lines result in the pair annihilations of the Majorana neutrinos to
the Majoron and the SM Higgs as shown in Figs. \ref{annihilation1}
and \ref{annihilation2} \footnote{Before the electroweak symmetry
breaking, the couplings of the SM leptons $\psi_L^{}$ and
$\ell_R^{}$ to the Majoron $\eta$ have no contributions to the
annihilations of the Majorana neutrinos $N$.}.

After an explicit calculations, we find the annihilation cross
sections to be
\begin{subequations}
\begin{eqnarray}
\label{cross1}
&&\sigma_{N_{i}^{}N_{i}^{}\rightarrow \eta\eta}^{}|\vec{v}|\nonumber\\
&=&\frac{1}{32\pi}\hat{h}_{i}^{4}\frac{1}{s}\left\{\frac{s^2_{}
(s-4M_{N_i^{}}^{2})-8M_{N_i^{}}^2s(s-M_\sigma^2)}{M_{N_i^{}}^2[(s-M_{\sigma}^{2})^{2}_{}+M_{\sigma}^{2}\Gamma_{\sigma}^{2}]}-4\right\}\nonumber\\
&&-\frac{1}{16\pi}\hat{h}_{i}^{4}\frac{1}{\sqrt{s(s-4M_{N_i^{}}^2)}}\left[1
+\frac{2s(s-M_\sigma^2)}{(s-M_\sigma^2)^2+M_\sigma^2\Gamma_\sigma^2}\right]\nonumber\\
&&\times
\ln\frac{\sqrt{s}-\sqrt{s-4M_{N_i^{}}^{2}}}{\sqrt{s}+\sqrt{s-4M_{N_i^{}}^{2}}}\,,\\
[5mm] \label{cross3}
&&\sigma_{N_{i}^{}N_{i}^{}\rightarrow \phi^\ast_{}\phi}^{}|\vec{v}|\nonumber\\
&=& \frac{1}{2\pi}\lambda_{3}^{2}
\frac{M^{2}_{N_i^{}}(s-4M_{N_i^{}}^{2})}{s[(s-M_{\sigma}^{2})^{2}_{}+M_{\sigma}^{2}\Gamma_{\sigma}^{2}]}\,,
\end{eqnarray}
\end{subequations} where
\begin{eqnarray}
|\vec{v}|=2\left(1-\frac{4M_{N_i^{}}^{2}}{s}\right)^{\frac{1}{2}}_{}\,,
\end{eqnarray}
is the relative velocity with $s$ being the squared center of mass
energy. The total cross section is then given by
\begin{eqnarray}
\sigma_{A}^{}|\vec{v}|= \sigma_{N_{i}^{}N_{i}^{}\rightarrow
\eta\eta}^{}|\vec{v}|+\sigma_{N_{i}^{}N_{i}^{}\rightarrow
\phi^\ast_{}\phi}^{}|\vec{v}|\,.
\end{eqnarray}

In order to determine the frozen temperature $T_{F}^{}$, at which
the annihilation of $N_{i}^{}$ becomes slower than the expansion of
the Universe, we should calculate the thermal average cross section,
\begin{eqnarray}
\langle\sigma_{A}^{}|\vec{v}|\rangle
=\frac{\displaystyle{\int_{4M_{N_i^{}}^{2}}^{\infty}s\sqrt{s-4M_{N_i^{}}^{2}}K_{1}^{}\left(\frac{\sqrt{s}}{T}\right)\sigma_{A}^{}|\vec{v}|
ds}}{\displaystyle{\int_{4M_{N_i^{}}^{2}}^{\infty}s\sqrt{s-4M_{N_i^{}}^{2}}K_{1}^{}\left(\frac{\sqrt{s}}{T}\right)
ds}}\,,
\end{eqnarray}
according to the standard manner \cite{luty1992,kw1980}, and then
require the reaction rate,
\begin{eqnarray}
\label{rate}
\Gamma=n_{N_{i}^{}}^{\textrm{eq}}\langle\sigma_{A}^{}|\vec{v}|\rangle
\end{eqnarray}
with the equilibrium number density
\begin{eqnarray}
n_{N_{i}^{}}^{\textrm{eq}}=\frac{1}{\pi^2}M_{N_i^{}}^{2}TK_2^{}\left(\frac{M_{N_i^{}}^{}}{T}\right)\,,
\end{eqnarray}
to be smaller the Hubble constant,
\begin{eqnarray}
\label{hubble}
H(T)=\left(\frac{8\pi^{3}_{}g_{\ast}^{}}{90}\right)^{\frac{1}{2}}_{}\frac{T^{2}_{}}{M_{\textrm{Pl}}^{}}
\end{eqnarray}
with the Planck mass $M_{\textrm{Pl}}^{}\simeq
10^{19}_{}\,\textrm{GeV}$ and the relativistic degrees of freedom
$g_{\ast}^{}\simeq 100$. From the equation,
\begin{eqnarray}
\label{departure} \Gamma \lesssim H(T)\,,
\end{eqnarray}
we can find a solution to $T_{F}^{}$ depending on $M_{N_i^{}}^{}$
and $f$.

For the purpose of analytical approximation, we expand the
annihilation cross section $\sigma_{A}^{}|\vec{v}|$ in powers of
$\displaystyle{\frac{1}{2}|\vec{v}|\leq 1}$ for calculating its
thermal average. Up to order $\mathcal{O}(|\vec{v}|^{2}_{})$, we
arrive at
\begin{eqnarray}
\label{thermal1}
\sigma_{A}^{}|\vec{v}|&\simeq&\frac{1}{8\pi}\left(\frac{1}{3}+\frac{\lambda_{3}^{2}}{4\lambda_{1}^{2}}
+\frac{5}{3\lambda_{1}^{}}\frac{M_{N_i^{}}^2}{f^2_{}}
\right)\frac{M_{N_i^{}}^{2}}{f^{4}_{}}|\vec{v}|^{2}_{}\,,
\end{eqnarray}
where the simplification, $s\ll M_{\sigma}^{2}$, has been adopted
since the larger $s$ has no significant contribution to the thermal
averaging in presence of the modified Bessel function
$\displaystyle{K_{1}^{}(\frac{\sqrt{s}}{T})}$. For
$\lambda_{1}^{}\simeq \lambda_{3}^{}=\mathcal{O}(1)$, the cross
section (\ref{thermal1}) can be simplified as
\begin{eqnarray}
\label{thermal2}
\sigma_{A}^{}|\vec{v}|&\simeq&\frac{1}{8\pi}\left(\frac{7}{12}+\frac{5}{3}\frac{M_{N_i^{}}^2}{f^2_{}}
\right)\frac{M_{N_i^{}}^{2}}{f^{4}_{}}|\vec{v}|^{2}_{}\,.
\end{eqnarray}
Since the leptogenesis usually occurs at $M_{N_i^{}}^{}\gtrsim
T_{lep}^{}> T_{sph}^{}\sim 100\,\textrm{GeV}$, we only need satisfy
the out-of-equilibrium condition (\ref{departure}) at $T=
M_{N_i^{}}^{}$, i.e.
\begin{eqnarray}
\label{departure2} \Gamma \lesssim
H(T)\left|_{T=M_{N_i^{}}^{}}^{}\right.\,.
\end{eqnarray}
Therefore we perform the equilibrium number density
\begin{eqnarray}
\label{number2}
n_{N_i^{}}^{\textrm{eq}}\left|_{T=M_{N_i^{}}^{}}^{}\right.&\simeq&0.16M_{N_i^{}}^{3}\,,
\end{eqnarray}
and the thermal averaging
\begin{eqnarray}
\label{thermal3}
\langle\sigma_{A}^{}|\vec{v}|\rangle\left|_{T=M_{N_i^{}}^{}}^{}\right.
\simeq 0.023\frac{M_{N_i^{}}^2}{f^{4}_{}}\,\,\, \textrm{with}\,\,
\langle|\vec{v}|^{2}_{}\rangle\left|_{T=M_{N_i^{}}^{}}^{}\right.\simeq
1\,.
\end{eqnarray}
Inputting (\ref{number2}) and (\ref{thermal3}) to (\ref{rate}) and
then to (\ref{departure2}) with (\ref{hubble}), we eventually obtain
the low bound on the breaking scale of the lepton number,
\begin{eqnarray}
\label{bound} f&\gtrsim& 0.1\times \left(M_{N_i^{}}^{3}
M_{\textrm{Pl}}^{}\right)^{\frac{1}{4}}_{}\nonumber\\
&\gtrsim& 0.1\times \left(T_{sph}^{3}
M_{\textrm{Pl}}^{}\right)^{\frac{1}{4}}_{}\nonumber\\
&\simeq&10^{5}_{}\,\textrm{GeV}\,.
\end{eqnarray}
In general, if we don't resort to the resonant effect
\cite{fps1995,fpsw1996,pilaftsis1997} by highly fine tuning,
$N_{i}^{}$ should be heavier than $10^{9}_{}\,\textrm{GeV}$
\cite{di2002,bdp2003} to induce a desired CP asymmetry so that the
lepton number should be broken at a much higher scale,
\begin{eqnarray}
\label{bound} f&\gtrsim&10^{10}_{}\,\textrm{GeV}\,.
\end{eqnarray}

Our conclusion may be applied to the singlet-triplet Majoron model,
where some triplet Higgs scalars $\xi$ with $L=-2$ are introduced to
the SM as well as the previously singlet Higgs scalar $\chi$.
Besides the quadratic and quartic interactions of these scalars,
there will be additional lepton number conserving Yukawa and quartic
interactions,
\begin{eqnarray}
\label{singlet-triplet} \mathcal{L}&\supset& -\frac{1}{2}
f_{ij}^{}~\overline{\psi^{c}_{L_i}}i\tau_{2}^{}\xi \psi_{L_j}^{} -
\kappa \chi^{\dagger}_{} \phi^{T}_{} i\tau_{2}^{}\xi
\phi\nonumber\\
&& -\lambda
 \left(\chi^\dagger_{} \chi\right)\textrm{Tr}\left(\xi^\dagger_{} \xi\right)\,.
\end{eqnarray}
The last term of the above Lagrangian shows that the singlet's VEV
will dominate the triplets' masses for certain range of the
parameters. In this case, similar with the pair annihilations of the
right-handed neutrinos in the singlet Majoron model, the scattering
processes of the triplets to the Majoron and to the SM Higgs will
also give us a low bound on the breaking scale of the lepton number.
Obviously, this constraint is not as strong as that in the singlet
Majoron model since the triplets' masses may be dominated by their
mass term.

Let us now consider a R-parity violating supersymmetric model of
thermal leptogenesis with right-handed neutrinos \cite{fv2006}. The
essential terms in the superpotential are
\begin{eqnarray}
\label{r-par} \mathcal{W}& = &\epsilon_{\alpha \beta}^{}
(Y_\nu^{})_{ij}^{} \hat N_i^{} \hat L^\alpha_j \hat H^\beta_u + {1
\over 2} M_{N_{ij}^{}}\hat
N_i^{} \hat N_j^{}\nonumber\\
&& + \epsilon_{\alpha \beta}^{} \lambda_i^{} \hat N_i^{} \hat
H^\alpha_d \hat H^\beta_u \,.
\end{eqnarray}
In the last term, lepton number is explicitly broken along with
R-parity violation. The VEVs of the Higgs doublets would also induce
a VEV to the scalar component of the right-handed neutrinos, which
is of the order of $(Y_\nu^{})_{ij}^{} \langle H_u^{} \rangle
\langle H_d^{} \rangle / M_N^{}$ and breaks the lepton number
spontaneously. As a result, there will be a would-be Majoron in the
model, which is very heavy. Unlike the previous discussions, these
would-be Majorons could have lepton number violating interactions,
that can constrain the scale of lepton number violation in this
model. Consider the term in the scalar potential $\left| {\partial
\mathcal{W} \over
\partial {H_u^{}}} \right|^2$, which gives rise to the scattering of
the Majorons $ \tilde N + \tilde N \to \tilde \nu H$. This
interaction will severely constrain the scale of leptogenesis in
this class of models.

In summary, we have studied the constraint on the Majoron models if
the leptogenesis is expected to explain the matter-antimatter
asymmetry of the Universe. Our calculations indicate that the
spontaneous breakdown of the global lepton number conservation in
the singlet Majoron model should happen at a high scale ($\gtrsim
10^{5}_{}\,\textrm{GeV}$) in order to guarantee the right-handed
neutrinos to be out of equilibrium before the sphaleron action
stops. This implies that some interesting phenomenology
\cite{jl1991} of the singlet Majoron model will be absent from the
LHC. We also point out that the similar constraint may exist in the
singlet-triplet Majoron model and the R-parity violating
supersymmetric Majoron model.

{\bf Acknowledgement}: We would like to thank Professor Ernest Ma
for his valuable comments.

\end{document}